\documentclass[a4paper,prd,twocolumn,aps,preprintnumbers,showpacs]{revtex4-1}

\usepackage{amsmath,amssymb}
\usepackage{epsfig}
\usepackage{graphicx}
\usepackage{dcolumn}
\usepackage{bm}

\def \be  {\begin{equation}}
\def \ee  {\end{equation}}
\def \beq  {\begin{equation}}
\def \eeq {\end{equation}}
\def \ba  {\begin{eqnarray}}
\def \ea  {\end{eqnarray}}
\def \baa {\begin{eqnarray*}}
\def \eaa {\end{eqnarray*}}
\def \bb  {\begin{thebibliography}}
\def \eb  {\end{thebibliography}}

\def \nn {\nonumber}

\def \lab #1 {\label{#1}}

\newcommand{\ie}{{\it i.e.}}

\newcommand\as{\alpha_s}
\newcommand\f{\frac}
\newcommand{\vect}[1]{\boldsymbol{#1}} 

\def \fracs #1#2 {\mbox{\small $\frac{#1}{#2}$}}

\def \bin #1#2 {{\left({#1}\atop{#2}\right)}}
\def\lapproxeq{{\ \lower 0.6ex \hbox{$\buildrel<\over\sim$}\ }}
\def\gapproxeq{{\ \lower 0.6ex \hbox{$\buildrel>\over\sim$}\ }}

\def\hepph  #1 {{hep-ph/#1 }}

\begin{document}

\title{Interplay of Threshold Resummation and Hadron Mass Corrections\\
in Deep Inelastic Processes}
\author{Alberto Accardi${}^{\,a,b}$, Daniele P. Anderle${}^{\,c}$, Felix Ringer${}^{\,c}$}
\affiliation{${}^{a}\,$Hampton University, Hampton, VA 23668, USA \\
${}^{b}\,$Jefferson Lab, Newport News, VA 23606, USA\\
${}^{c}\,$T\"ubingen University, 72076 T\"ubingen, Germany}
\begin{abstract}
We discuss hadron mass corrections and threshold resummation for deep-inelastic scattering $\ell N\rightarrow\ell' X$ and semi-inclusive annihilation $e^+e^-\rightarrow h X$ processes, and provide a prescription how to consistently combine these two corrections respecting all kinematic thresholds. We find an interesting interplay between threshold resummation and target mass corrections for deep-inelastic scattering at large values of Bjorken $x_B$.
In semi-inclusive annihilation, on the contrary, the two considered corrections are relevant in different kinematic regions and do not affect each other. A detailed analysis is nonetheless of interest in the light of recent high precision data from BaBar and Belle on pion and kaon production, with which we compare our calculations.
For both deep inelastic scattering and single inclusive annihilation, the size of the combined corrections compared to the precision of world data is shown to be large. Therefore, we conclude that these theoretical corrections are relevant for global QCD fits in order to extract precise parton distributions at large Bjorken $x_B$, and fragmentation functions over the whole kinematic range.

\end{abstract}

\date{\today}
\maketitle

\section{Introduction \label{intro}}

Predictions from QCD rely on perturbative calculations of parton-level hard scattering processes as well as on non-perturbative input in the form of parton distribution functions (PDFs) and fragmentation functions (FFs). On the one hand, PDFs contain information about the distributions of quarks and gluons in hadrons, which is relevant for processes with initial-state hadrons. On the other hand, FFs describe the fragmentation of an outgoing parton into the observed hadron and, to some extent, may be viewed as the final-state analogue of PDFs. The applicability of this framework within perturbative QCD was established in factorization theorems~\cite{Collins:1989gx} allowing one to absorb long-distance dynamics into these two universal non-perturbative objects. Therefore, the predictive power of QCD relies crucially on the precise knowledge of PDFs and FFs, that are nowadays extracted from a global analysis of a wide set of experimental data, see Refs.~\cite{Jimenez-Delgado:2013sma,Forte:2013wc,Albino:2008gy} for recent reviews. 

Modern PDF fits~\cite{Martin:2009iq,Owens:2012bv,Lai:2010vv,*Ball:2014uwa} are available within a next-to-leading order (NLO) framework and most of them also at (partial) next-to-next-to-leading order. Key data sets for the extraction of PDFs are provided by measurements of inclusive deep-inelastic scattering (DIS) $\ell N\rightarrow \ell'X$, which is one of the two processes that we are considering in this work. Despite a lot of progress in the past years, large uncertainties are still present for large values of the parton momentum fraction $x$ \cite{Accardi:2013pra}. As it turns out, it is precisely this region that is particularly relevant at the LHC, when trying to find signals of new physics in, for example, (di-)jet measurements~\cite{Chatrchyan:2012bja,*Aad:2011fc}. Furthermore, the large-$x$ region is also interesting as it can provide a window into the non-perturbative dynamics of the color confinement mechanism holding quarks and gluons inside hadrons \cite{Holt:2010vj,Courtoy:2014xea}.

On the experimental side, improvements for the gluon PDF at large-$x$ can be obtained from jet data taken at the Tevatron and the LHC, direct photon production in fixed target experiments, and from longitudinal DIS structure functions. Concerning quark PDFs, the present focus is mostly on low-energy experiments carried out for example at JLab \cite{Dudek:2012vr}, with important information coming from directly reconstructed $W$ charge asymmetries at the Tevatron \cite{Accardi:2013pra}. On the theoretical side a number of corrections to the pQCD calculations of these events are needed in order to harvest fully the available and upcoming experimental data, and extract precise large-x quark and gluon PDFs from global QCD fits. These corrections include, in particular, resummation of threshold logarithms, Target Mass Corrections (TMCs),  higher-twist diagrams, and nuclear corrections when nuclear targets are considered. The last three have been included consistently, for example,  in the CTEQ-JLab collaboration PDF fits \cite{Owens:2012bv} and the fits by Alekhin and collaborators \cite{Alekhin:2012ig}, allowing to substantially extend the range in $x$ of the fitted DIS data. Threshold resummation has been considered in the past to estimate the theoretical errors of PDFs or was used for fits of only a subset of the data  \cite{Sterman:2000pu,Martin:2003sk,Sato-PhDthesis}, but has not yet been fully included for all relevant data sets in a global QCD fit.

In the first part of this work, we consider the interplay between two major corrections to the standard NLO formalism for DIS both of which have their greatest impact at large-$x$, namely TMCs and higher order contributions derived from threshold resummation. Here we choose the collinear factorization TMC framework of Accardi and Qiu \cite{Accardi2009}, that contrary to most other approaches \cite{Georgi:1976ve,*Schienbein:2007gr} respects the kinematic $x_B \leq 1$ bound on the Bjorken variable.
Threshold resummation for QCD processes was derived in~\cite{Sterman:1981jc,*Gatheral:1983cz,*Frenkel:1984pz,*Catani:1989ne} and recently revisited in \cite{Anderle:2012rq,Anderle:2013lka,Anderle:2013pla}. Large logarithms that need to be resummed to all orders arise near the phase space boundary where gluon radiation is limited. We perform the resummation at the level of next-to-leading logarithmic (NLL) accuracy. In particular, we derive a resummation procedure that also respects the Bjorken $x_B$ bound when a non-zero target mass is considered, and can therefore be consistently combined with the TMC calculation. 
We discuss the interplay of both kinds of corrections, and assess their relevance for PDF global fits by comparing them to a selection of world data on DIS scattering. 

Unlike for PDFs, global fits of FFs \cite{deFlorian:2007aj, deFlorian:2014xna, Albino:2008fy, Epele:2012vg,*Hirai:2007cx,*Soleymaninia:2013cxa} are less constrained by presently existing data sets.
One of the main sources of constraints on FFs is semi-inclusive annihilation (SIA) $e^+e^-\rightarrow hX$ which we are going to consider in this work. Recently, very precise data sets from BELLE~\cite{Leitgab:2013qh} and BaBar~\cite{Lees:2013rqd} became available, where the statistical accuracy is partially in the sub one percent level. In addition, a very fine binning was applied over a wide range of the fragmentation variable $x_E=2E^h/\sqrt{s}$ reaching up to $\approx 0.95$. Here, $E^h$ is the energy of the observed hadron in the center-of-mass system (c.m.s.) and $\sqrt{s}=10.5$~GeV is the energy for collisions at both experiments. This offers a new possibility for studying effects that go beyond the standard NLO framework and for learning more about QCD dynamics in fragmentation processes. On the theory side, the present day state of the art is NLO in QCD. Several additional effects, including small-$x$ resummation, threshold resummation and hadron mass effects have been studied in~\cite{Albino:2008fy,Albino:2005gd}. 

In the second part of this paper we revisit calculations of hadron mass corrections (HMCs) and threshold resummation in analogy to our DIS analysis. We present for the first time (to our knowledge) a resummed calculation for kaon SIA events, and compare our kaon and pion production cross section to the recent BELLE and BaBar data.
In contrast to the Operator Product Expansion based formalism for mass corrections, the approach in~\cite{Accardi2009} may be generalized to other processes, such as semi-inclusive deep-inelastic scattering~\cite{Accardi:2009md}. Here we extend this framework to SIA in electron-positron scattering, we perform a detailed analysis of the effects of the produced hadron mass on the parton-level kinematics, and evaluate their numerical consequences. We note that previous studies of HMCs were carried out in~\cite{Albino:2008fy,Albino:2005gd}, showing in particular that inclusion of these HMCs in global FF fits results in better $\chi^2$ values. We then consider the combination with threshold resummation~\cite{Cacciari:2001cw} in the framework of the so-called ``crossed resummation''~\cite{Sterman:2006hu}, that exploits similarities between various color-singlet QCD processes such as DIS, SIA, Drell-Yan, and semi-inclusive deep-inelastic scattering. We will again also build upon the recent threshold resummation studies in  \cite{Anderle:2012rq,Anderle:2013lka,Anderle:2013pla}.
Contrary to DIS, we find that HMCs are dominant at low $x_E$, whereas threshold resummation is again most relevant for large $x_E$. We analyze the crosstalk of these effects and evaluate their relevance to global FF fits by comparing these to the new data sets from BELLE and BaBar. 

Our paper is organized as follows. In Sec.~\ref{sec:tmc}, we discuss TMCs and threshold resummation in DIS before we derive our prescription to combine both. In order for our paper to be self-contained, we briefly review the TMC derivation of  \cite{Accardi2009}, and provide some basic formulas concerning threshold resummation in order to establish our notation. Then, we analyze the numerical relevance of the corrections, and compare these to a selection of world DIS data. In Sec.~\ref{sec:hmc}, we discuss SIA following the same steps as for the DIS case before, and compare our numerical results to the recent BELLE and BaBar data on pion and kaon production. Finally, we draw our conclusions in Sec.~\ref{sec:conclusions}.

\section{Target Mass Corrections and Resummation for DIS \label{sec:tmc}}
 
\subsection{Target Mass Corrections \label{ssec:tmc1}}

\begin{figure}[t]
\vspace*{-2mm}
\hspace*{0cm}
\includegraphics[width=0.35\textwidth]{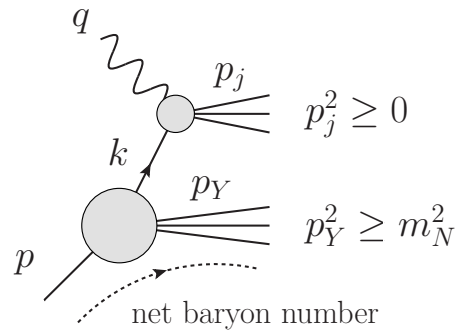}
\caption{\sf Diagram for DIS specifying all momenta. The net baryon number is shown to flow into the target jet. Figure taken from~\cite{Accardi2009}. \label{fig:dis1}}
\end{figure}

In DIS, a parton of momentum $k$ belonging to a nucleon of momentum $p$ is struck by a virtual photon of momentum $q$. This generates in the final state a target jet with momentum $p_Y$ and a current jet with momentum $p_j$, see Fig.~\ref{fig:dis1}. We work in a ``collinear'' frame where the spacial components of $p$ and $q$ are parallel and directed along the longitudinal axis, and we parametrize the involved the momenta $p,q,k$ following~\cite{Accardi2009}:
\ba\label{eq:momentadis}
p^\mu & = & p^+\bar n^\mu+\frac{m_N^2}{2p^+} n^\mu \, ,\nn \\
q^\mu & = & -\xi p^+\bar n^\mu+\frac{Q^2}{2\xi p^+} n^\mu \, ,\nn \\
k^\mu & = &  xp^+\bar n^\mu+\frac{k^2+k_T^2}{2xp^+}n^\mu+\vect{k}_T^\mu\, ,
\ea
In this expression $p^+$ can be regarded as a parameter for boosts along the longitudinal axis. The light-cone vectors $n^\mu$ and $\bar n^\mu$ satisfy
\be
n^2=\bar n^2=0\,\quad n\cdot\bar n=1\, ,
\ee
and the plus- and minus- components of a general 4-vector $a$ are given by
\be
a^+=a\cdot n \quad a^-=a\cdot \bar{n}.
\ee
The momenta are parametrized in terms of the external (\ie, experimentally measurable) variables
\be
x_B=\frac{-q^2}{2p\cdot q},\quad Q^2=-q^2,\quad p^2=m_N^2\, ,
\ee
where $m_N$ is the target mass and $Q^2$ the photon virtuality. 
The parton fractional light-cone momentum with respect to the nucleon is a kinematic internal (\ie, non measurable) variable and is defined by
\begin{equation}
\label{eq:frac_mom}
x=\f{k^+}{p^+}.
\end{equation}
In an analogous way we can define the virtual boson fractional momentum as
\begin{equation}
\label{eq:Nacht}
\xi=-\f{q^+}{p^+}=\f{2 x_B}{1+\sqrt{1+4 x_B^2 m^2_N /Q^2}},
\end{equation}
which is an external kinematic variable and coincides with the Nachtmann variable~\cite{Nachtmann:1973mr}. 
The target's mass can be neglected in the Bjorken limit $Q^2 \gg m_N^2$ at fixed $x_B$, and in many analysis is omitted from the outset. (This is fine for unpolarized scattering but poses problems for the definition of the nucleon's spin in the case of polarized scattering.) In this paper we explicitly work at finite $Q^2$ and verify that our result correctly reproduces the ``massless target'' formulas in the Bjorken limit, where $\xi\rightarrow x_B$. \\

In order to perform collinear factorization of the DIS structure functions, one expands the momentum $k$ of the struck parton around its positive light-cone component $xp^+\bar n^\mu$, and neglects in the kinematics the transverse components, as well as (for light quarks) the negative light-cone component. This is equivalent to kinematically treating the parton as massless and collinear to the parent nucleon from the very beginning, and setting $k^2=0$ and ${\bm k_T}^\mu=0$ in Eq.~\eqref{eq:momentadis} from the beginning. A more detailed analysis of the collinear kinematics keeping $k^2\neq 0$ is left for future work~\cite{Accardi-Anderle-Ringer-2}. \\

The parton's momentum fraction $x$ then appears as an integration variable in the structure functions, that are given by a convolution integral of perturbatively calculable coefficient functions and non-perturbative PDFs \cite{Collins:1989gx,Qiu:1988dn}. Following~\cite{Accardi2009}, we may derive limits on the $dx$ integration by examining both the external and internal kinematics of the diagram shown in Fig.~\ref{fig:dis1}, and apply four momentum and net baryon number conservation. This latter, in particular requires that at least one baryon of mass larger than $m_N$ be present in the final state. This can appear in either the target jet (lower right part in Fig.~\ref{fig:dis1}) or the current jet (upper right part in Fig.~\ref{fig:dis1}). Unless the rapidity difference between the current jet and the target jet is too small \cite{Berger:1987zu,*Mulders:2000jt}, the baryon mass appears in the latter~\cite{Accardi2009}, so that $p_Y^2\geq m_N^2$ and $p_j\geq 0$. Next, considering four-momentum conservation the hard-scattering vertex we find
\begin{align}
\label{eq:firstdisconstraint}
0 \leq p_j^2 & = (q+k)^2 = \left(1-\frac{\xi}{x} \right)\frac{Q^2x}{\xi}\, ,
\end{align}
where we used the momenta defined in Eq.~(\ref{eq:momentadis}). In order to obtain another constraint on the $dx$ integration, it is not sufficient to analyze the other vertex. Instead, we have to consider the invariant momentum squared of the whole process:
\begin{align}\label{eq:seconddisconstraint1}
(q+p)^2 & = (p_j+p_Y)^2 \geq (q+k)^2+m_N^2 \, .
\end{align}
where we used $p_Y^2 \geq m_N$, as previously discussed, as well as $2p_j\cdot p_Y\geq 0$ since both final state jets consist of on-shell particles. 
Evaluating $(q+k)^2$ as before in Eq.~(\ref{eq:firstdisconstraint}) we finally obtain
\be
1-\frac{1}{x_B}\leq \left(1-\frac{x}{\xi} \right) .
\ee
In summary, the parton's fractional momentum $x$ is kinematically bound by
\be\label{eq:limitsDIS}
\xi\leq x\leq \xi/x_B \, ,
\ee
as first discussed in ~\cite{Accardi2009}.

The structure functions including the finite target mass kinematics 
we have discussed can be written in collinear factorization at leading twist as~\cite{Accardi2009, Brady:2011uy}
\begin{align}
\label{eq:Ftmc}
{\cal F}_1^{\text{TMC}}(x_B,Q^2) & \equiv 2 F_1^{\text{TMC}}(x_B,Q^2) = {\cal F}_1(\xi,Q^2) \, ,\nn \\
{\cal F}_2^{\text{TMC}}(x_B,Q^2) & \equiv \frac{1}{x_B}F_2^{\text{TMC}}(x_B,Q^2) = \frac{1}{\rho^2} {\cal F}_2(\xi,Q^2)\, ,\nn  \\
{\cal F}_L^{\text{TMC}}(x_B,Q^2) & \equiv \frac{1}{x_B} F_L^{\text{TMC}}(x_B,Q^2)={\cal F}_L(\xi,Q^2) \, ,
\end{align}
where for convenience we defined
\be
\label{eq:rho}
\rho^2=1+\frac{4x_B^2m_N^2}{Q^2}\, \ .
\ee
Note that we adopted the convention of Ref.~\cite{Anderle:2012rq,deFlorian:2012wk} for the ${\cal F}_i^{\text{TMC}}$ structure functions in terms of the customary ones appearing in the Lorentz decomposition of the hadronic tensor satisfying ${\cal F}_L^{\text{TMC}}(x_B,Q^2)=\rho^2 {\cal F}_2^{\text{TMC}}(x_B,Q^2)-{\cal F}_1^{\text{TMC}}(x_B,Q^2)$.
On the right hand side of Eq.~(\ref{eq:Ftmc}), convolution integrals appear
\be\label{eq:FTMC}
{\cal F}_i(\xi,Q^2)=\sum_f\int_\xi^{\xi/x_B}\frac{dx}{x}f(x,\mu^2)\,{\cal C}_f^i\left(\frac{\xi}{x},\frac{Q^2}{\mu^2},\as(\mu^2)\right) \ ,
\ee
where the integration over $dx$ ranges only over the region allowed by the limits in Eq.~(\ref{eq:limitsDIS}). The notation we use implies that whenever the lower limit exceeds the upper limit the integral is zero, so that the structure functions are indeed zero in the kinematically forbidden region $x_B>1$. The functions $f(x,\mu^2)$ denote the distribution of a parton of flavor $f$ in the target nucleon and the sum runs over $f=q,\bar q,g$, with $q$ a shorthand for all active quark flavors. The hard-scattering coefficient functions ${\cal C}_f^i$ encode the short-distance hard scattering of the virtual photons with partons from the nucleon target, and are independent of the mass of the latter. They can be calculated in perturbative QCD order-by-order in powers of the strong coupling constant,
\be
{\cal C}_f^i=C^{i,(0)}_f+\frac{\as(\mu^2)}{2\pi}C_f^{i,(1)}+{\cal O}(\as^2) \, ,
\ee
which are related by ${\cal C}_f^L = {\cal C}_f^2 - {\cal C}_f^1$ for massless partons. 
For example, at leading-order (LO) we have
\begin{align}
C^{2,(0)}_{q,\bar q}(\hat x)&=e_{q}^2\delta(1-\hat x),\; C^{2,(0)}_{g}(\hat x)=0 \, ,\nn \\
C^{L,(0)}_{q, \bar q, g}(\hat x)&=0 \, ,
\end{align}
with $\hat x=\xi/x$. $F^{\mathrm{TMC}}_2$ reduces to the target mass corrected version of the parton model \cite{Kretzer:2003iu,Aivazis:1993kh} except for a step function imposing the proper kinematic bounds:
\be
F^{\mathrm{TMC}}_2(x_B,Q^2)=\frac{x_B}{\rho^2}\sum_{f=q,\bar q} e_f^2 f(\xi,Q^2) \theta(1-x_B)\, .
\ee
For completeness, we list all the relevant coefficient functions ${\cal C}^{i}_f$ up to NLO in Appendix \ref{ap:DIScoefFunc}. \\

\begin{figure}[t]
\hspace*{-.5cm}
\includegraphics[width=0.45\textwidth]{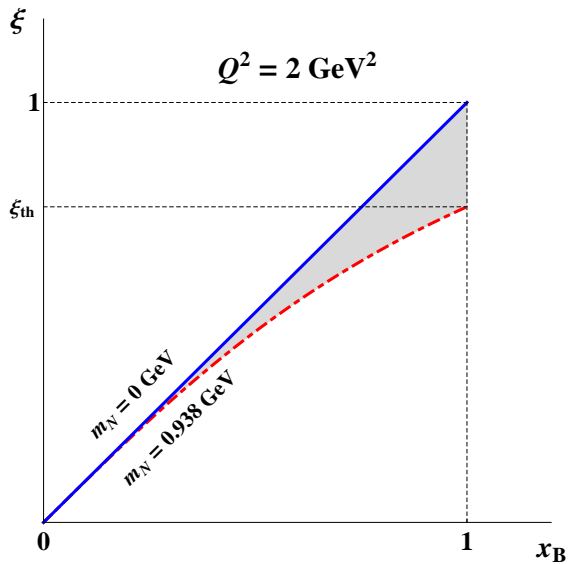}
\caption{\sf The Nachtmann variable $\xi$ as a function of $x_B$ with $m_N=0$ GeV (solid blue line) and $m_N=0.938$ GeV (dash-dotted red line) at $Q^2=2~\mathrm{GeV}^2$. \label{fig:xixB}}
\end{figure}

Note that in the large $Q^2$ limit (in which $M^2/Q^2 \rightarrow 0$), as well as in the small Bjorken-x limit $x_B \rightarrow 0$, the Nachtmann variable $\xi \rightarrow x_B$ and $\rho \rightarrow 1$, so that  
\begin{align}
  {\cal F}_i^{\text{TMC}}(x_B,Q^2) \rightarrow {\cal F}_i(x_B,Q^2)
\end{align}
and the usual massless target formulas are recovered. Conversely, in the $x_B\rightarrow 1$ limit, the Nachtmann variable $\xi \rightarrow \xi_{\mathrm{th}}$, where 
\begin{align}
\xi_{\mathrm{th}} =  \frac{2}{1+\sqrt{1+4 m_N^2/Q^2}} \, ,
\end{align}
and $\xi$ differs maximally from $x_B$ (see Fig.~\ref{fig:xixB}). Therefore, in this limit, TMC effects are the largest. Since the integral over $dx$ is limited to the region defined by the kinematic bounds in Eq.~(\ref{eq:limitsDIS}), the structure functions ${\cal F}_i^{\text{TMC}}$ have support only over the physical region at $x_B\leq 1$. This is the defining characteristics of the treatment of TMCs proposed in Ref.~\cite{Accardi2009} and sets this apart from most other TMC prescriptions, that in fact violate that bound and allow for non-zero structure functions also at $x_B>1$. When combining TMCs with threshold resummation in Section~\ref{ssec:tmc22}, we will pay special attention to preserve this feature of our TMC treatment and not introduce a spurious violation of the Bjorken-$x$ bound.

\subsection{Threshold Resummation for DIS \label{ssec:tmc21}}

The DIS coefficient functions ${\cal C}_f^i$ contain singular distributions. Near threshold they can get large and weaken or even violate the convergence of the perturbative expansion in the strong coupling constant. Therefore, they have to be taken into account to all orders via threshold resummation. At NLO, singular distributions only appear in the structure function ${\cal F}_1$ (or equivalently ${\cal F}_2$) but not in ${\cal F}_L$. In the $\overline{\mathrm{MS}}$ scheme, they read
\ba\label{eq:NLOth}
C^{1,(1)}_{q,\mathrm{th}}(x) & = & C_F\left[(1+x^2)\left(\f{\ln(1-x)}{1-x}\right)_+ -\f{3}{2}\f{1}{(1-x)}_+\right. \nn \\
&&\qquad\;\left. -\left(\f{9}{2}+\f{\pi^2}{3}\right)\delta(1-x) \right] \, ,
\ea
where the plus-distribution is defined as
\be\label{eq:plusdist}
\int_0^1 dx\, f(x)[g(x)]_+ \equiv
\int_0^1dx \left(f(x)-f(1)\right) g(x)
\ee
In general, at a given order $k$ in the perturbative expansion, the coefficient function contains logarithms of the form
\be
\as^k\left(\f{\ln^n(1-x)}{1-x}\right)_+,\quad\mathrm{with}\;n\leq 2k-1\, .
\ee
Performing the resummation at NLL, we fully take into account contributions down to $n=2k-3$ at all orders. In other words, resummation at NLL accuracy sums up correctly the three most dominant towers of threshold logarithms. Results at next-to-next-to leading logarithmic accuracy were derived in~\cite{Blumlein:2006pj,Moch:2009my}, where the next two subleading towers of threshold logarithms are also correctly taken into account. However, the main phenomenological effects are already captured at the level of NLL. In addition, the proposed prescription for combining TMC and resummation, as discussed in the next subsection, is independent of the accuracy of resummation that we are considering. \\

In the massless limit, $m_N^2/Q^2 \rightarrow 0$, resummation may be performed by introducing Mellin moments in $x_B$ of the massless structure functions:
\begin{align}
{\cal F}_{1}^N(Q^2) & = \int dx_B\, x_B^{N-1}\, {\cal F}_1(x_B,Q^2)\nn\\
& = {\cal C}_f^{1,N}\!\left(Q^2/\mu^2,\as(\mu^2)\right)\cdot f^N(\mu^2)\, ,
\end{align}
where
\begin{align}
{\cal C}_f^{1,N} & = \left(\int_0^1dx\, x^{N-1}{\cal C}_f^1\left(x,Q^2/\mu^2,\as(\mu^2)\right)\right) \\
f^N & = \left(\int_0^1 dy\, y^{N-1} f(y,\mu^2) \right) \,
\end{align}
and the superscript $N$ denotes the dependence on the complex Mellin variable $N$. The Mellin space expression of the NLO coefficient function up to terms that are suppressed as ${\cal O}(1/N)$ and choosing $\mu^2=Q^2$ is given by
\be
C^{1,(1),N}_{q}=C_F\left[\ln^2\bar N+\f{3}{2}\ln\bar N-\f{9}{2}-\f{\pi^2}{6}\right]\, ,
\ee
where large logarithms in $\bar N=Ne^{\gamma_E}$ correspond to large logarithms in $1-x$ in Eq.~(\ref{eq:NLOth}). The resummed DIS coefficient function for the structure function ${\cal F}_1$ reads to NLL~\cite{Sterman:2006hu,Cacciari:2001cw,Anderle:2012rq,Catani:1989ne}:
\ba\label{resummed1}
 {\cal{C}}^{1,N}_{q,\mathrm{res}}(Q^2/\mu^2,\as(\mu^2))&=&e_q^2 H_q\left(Q^2/\mu^2,\alpha_s(\mu^2)\right)\nn\\[2mm]
&&\hspace{-3cm} \times \Delta_q^N(Q^2/\mu^2,\as(\mu^2))\, J_q^N(Q^2/\mu^2,\as(\mu^2)) \, .
\ea
The radiative factor $\Delta^N_q$ describes gluon radiation from the initial quark that is both soft {\it and} collinear. The function $J_q^N$ takes into account collinear ({\it i.e.} soft and hard) emissions from the unobserved parton in the final state. The two functions $\Delta^N_q$ and $J_q^N$ are given by the following two exponentials
\begin{align}\label{eq:deltaJ}
\log \Delta_q^N &\equiv \int_0^1 dx \frac{x^N -1}{1-x}
  \int_{Q^2}^{(1-x)^2 Q^2} \frac{dk_\perp^2}{k_\perp^2} 
  A_q(\alpha_s(k_\perp^2)),\nn\\
\log J_q^N &\equiv \int_0^1 dx \frac{x^N -1}{1-x}\bigg\{
  \int_{(1-x)^2Q^2}^{(1-x) Q^2} \frac{dk_\perp^2}{k_\perp^2} 
  A_q(\alpha_s(k_\perp^2)) \nn\\[2mm]
& \hspace*{4mm}+\frac{1}{2}B_q(\alpha_s((1-x)Q^2)\bigg\} \, .
\end{align}
The functions $A_q(\alpha_s)$ and $B_q(\alpha_s)$ can be calculated in a perturbative expansion in the strong coupling constant
\ba
A_q(\alpha_s) & = & \frac{\alpha_s}{\pi} A_q^{(1)} + \left(\frac{\alpha_s}{\pi}\right)^2 A_q^{(2)} + \dots\;, \nn \\
B_q(\alpha_s) & = & \frac{\alpha_s}{\pi} B_q^{(1)} + \left(\frac{\alpha_s}{\pi}\right)^2 B_q^{(2)} +\ldots \; . 
\ea
The relevant coefficients for resummation at NLL accuracy are given by 
\ba
 A_q^{(1)} & = & C_F, \quad A_q^{(2)} = \frac{1}{2} C_F\left[C_A
 \left(\frac{67}{18}-\frac{\pi^2}{6}\right) - \frac{5}{9} N_f\right], \nn \\
B_q^{(1)} & = & -\frac{3}{2} C_F,
\ea
where $C_F=4/3$, $C_A=3$ and $N_f$ is the number of active flavors. Finally, the hard-scattering coefficient $H_q$ in Eq.~(\ref{resummed1}) reads 
\begin{align}\label{Hq}
H_q & \left(Q^2/\mu^2,\alpha_s(\mu^2)\right) \\
& =1 + \frac{\alpha_s}{2\pi} 
C_F \left(-\frac{9}{2} - \frac{\pi^2}{6} +
\frac{3}{2}\ln \frac{Q^2}{\mu^2}\right)\nn
+{\cal O}(\alpha_s^2).
\end{align}
Expanding the exponents in Eq.~(\ref{eq:deltaJ}) up to NLL accuracy, we find~\cite{Vogt:2000ci,Gardi:2002xm,Corcella:2005us,Grunberg:2009vs,Schaefer:2001uh,Anderle:2012rq}
\ba
\log\Delta_q^N\, & = &\, \ln \bar N h_q^{(1)}(\lambda)+  h_q^{(2)} \left(\lambda, \frac{Q^2}{\mu^2},\frac{Q^2}{\mu_F^2}\right) \nn\\
\log J_q^N&=&\ln \bar{N} f_q^{(1)}(\lambda) +f_q^{(2)}\left(\lambda, \frac{Q^2}{\mu^2}\right) \,,
\ea
where $\lambda=b_0\as(\mu^2)\ln\bar N$. The two functions $h_q^{(1)},\,f_q^{(1)}$ ($h_q^{(2)},\,f_q^{(2)}$) collect all leading logarithmic (next-to-leading logarithmic) contributions in the exponent of the type $\as^k\ln^n\bar N$ with $n=k+1$ ($n=k+2$). The functions $h^{(1)}$ and $h^{(2)}$ for $\Delta_q^N$ are given by
\begin{align}\label{eq:h}
h_q^{(1)}&(\lambda) =  \frac{A_q^{(1)}}{2 \pi b_0 \lambda}\left[2\lambda + 
(1- 2\lambda) \ln(1-2\lambda)\right], \nn \\[2mm]
h_q^{(2)} & \left(\lambda, \frac{Q^2}{\mu^2},\frac{Q^2}{\mu_F^2}\right) =  -
\frac{A_q^{(2)}}{2 \pi^2 b_0^2} \left[2\lambda + \ln (1-2\lambda)\right] \\[2mm]
& +\frac{A_q^{(1)}b_1}{2\pi b_0^3}\left[2\lambda + 
\ln(1-2\lambda)+ \frac{1}{2} \ln^2(1-2\lambda)\right] \nonumber  \\[2mm]
& + \frac{A_q^{(1)}}{2\pi b_0} \left[2 \lambda + 
\ln(1-2\lambda)\right] \ln\frac{Q^2}{\mu^2}  - \frac{A_q^{(1)}}{\pi b_0} \lambda
 \ln \frac{Q^2}{\mu_F^2}. \nn
\end{align}
The functions $f^{(1)}$ and $f^{(2)}$ for $J_q^N$ are given by
\ba
f_q^{(1)}(\lambda) &=& h_q^{(1)}\left(\frac{\lambda}{2}\right)- h_q^{(1)}(\lambda), \nonumber \\[2mm]
f_q^{(2)}\left(\lambda,\frac{Q^2}{\mu^2}\right) &=& 
2 h_q^{(2)}\left(\frac{\lambda}{2}, \frac{Q^2}{\mu^2},1\right)- 
h_q^{(2)}\left(\lambda, \frac{Q^2}{\mu^2},1\right) \nn \\
&& +\frac{B_q^{(1)}}{2\pi b_0}\ln(1-\lambda).
\ea
Here, the $b_0,\,b_1$ are the coefficients of the QCD beta function.\\

In the end, we go back to $x$-space by numerically performing the Mellin inverse, which is given by
\begin{align}\label{eq:mellininverse0}
{\cal F}_{1,\mathrm{res}}(x_B,Q^2) & =\int_{{\cal C}_N}\frac{dN}{2\pi i}\,x_B^{-N} \nn \\
& \times \,{\cal C}^{1,N}_{q,\mathrm{res}}(Q^2/\mu^2,\as(\mu^2))\,f^N(\mu^2)\, .
\end{align}
The contour ${\cal C}_N$ is taken to run between the rightmost pole of the moments of the PDFs and the Landau pole following the minimal prescription of~\cite{Catani:1996yz}. After the Mellin inverse is taken, we match to the full NLO which is still a good approximation away from threshold. We avoid double counting of threshold distributions at NLO by considering the matched combination
\be\label{eq:matching}
{\cal F}_{\mathrm{match}}={\cal F}_{\mathrm{res}}-\left.{\cal F}_{\mathrm{res}}\right|_{{\cal O}(\alpha_s)}+{\cal F}_{\mathrm{NLO}} \, .
\ee

\begin{figure*}[t!]
\vspace*{-1cm}
\centering
\includegraphics[width=\textwidth,clip]{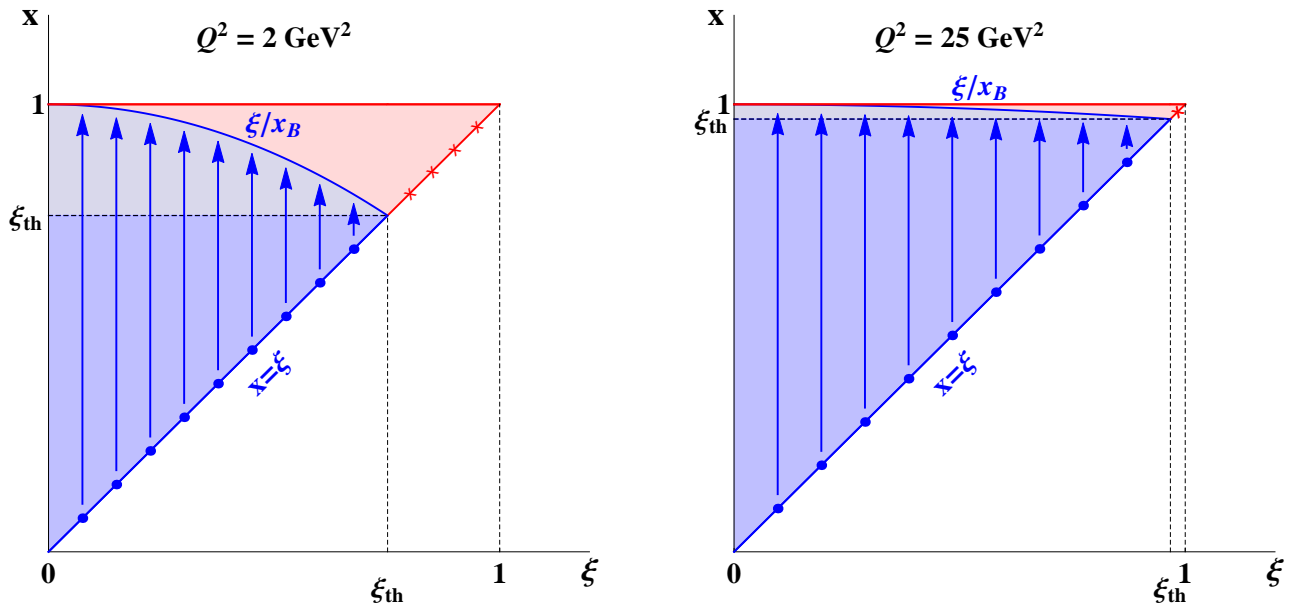}
\vspace*{-.8cm}
\caption{\sf On the left (right) hand side the integration regions for $Q^2=2\;\mathrm{GeV}^2$ ($Q^2=25\;\mathrm{GeV}^2$) concerning Eq.~(\ref{eq:DIS3}) are shown. The blue dots denote the boundary where the threshold singularities arise and the arrows indicate the direction of integration. \label{fig:integration1}}
\end{figure*}

\subsection{Combining TMC and Threshold Resummation \label{ssec:tmc22}}

After target mass corrections, the integration over the parton's momentum fraction in the collinear factorization formula \eqref{eq:FTMC} ranges from $\xi$ to $\xi/x_B$. As a consequence, the Mellin moments of the structure function are no longer the product of the moments of the coefficient function ${\cal C}_1$ and the parton distribution $f$. One therefore may be tempted to express the structure function \eqref{eq:FTMC} as
\begin{align}
\label{eq:DIS2}
{\cal F}^{\text{TMC}}_1 = \int_\xi^1\frac{dx}{x}{\cal C}_f^1
  \left( \f{\xi}{x}\right)f(x) 
  - \int_{\xi/x_B}^{1}\frac{dx}{x}{\cal C}_f^1\left(\f{\xi}{x}
  \right)f(x)   \, ,
\end{align}
where for ease of notation we omitted any dependence of the coefficient functions and the PDFs on the scale $Q^2/\mu^2$ and on $\as(\mu^2)$. The advantage of this reformulation is that the first term is integrated up to 1 (and differs from the Bjorken limit approximation only by a $x_B \rightarrow \xi$ replacement), so that its Mellin transform would indeed be given by the product of moments of the coefficient and parton distribution functions. However, written in this way, ${\cal F}^{\text{TMC}}_1$ acquires support also in the unphysical region $x_B>1$, where it actually becomes negative after crossing 0 at $x_B=1$.

A better way to manipulate the structure function convolution in  Eq.~\eqref{eq:FTMC} in order to obtain a product of moments after performing its Mellin transformation, is to write 
\begin{align}
\label{eq:DIS3}
{\cal F}^{\text{TMC}}_1 
  = \int_\xi^{\xi_{\mathrm{th}}}\frac{dx}{x} {\cal C}_f^1 
  \left( \f{\xi}{x} \right) f(x) 
  + \int_{\xi_{\mathrm{th}}}^{\xi/x_B}\frac{dx}{x}{\cal C}_f^1 
  \left(\f{\xi}{x}\right) f(x) \ .
\end{align}
In the small $x_B$ limit only the first term on the right hand side survives, and the massless limit is recovered, as it should be. In the $x_B\rightarrow 1$ limit, each term separately tends to zero and remain zero for larger values of $x_B$. Therefore, the structure function as well remains zero in the unphysical region $x_B>1$, as it happens with the original Eq.~\eqref{eq:FTMC}. This is then a good starting point for performing the resummation of threshold distributions in a way that respects the partonic and hadronic kinematics discussed in Section~\ref{ssec:tmc1}.

\begin{figure}[t]
\vspace*{-2mm}
\hspace*{-.5cm}
\includegraphics[width=0.5\textwidth]{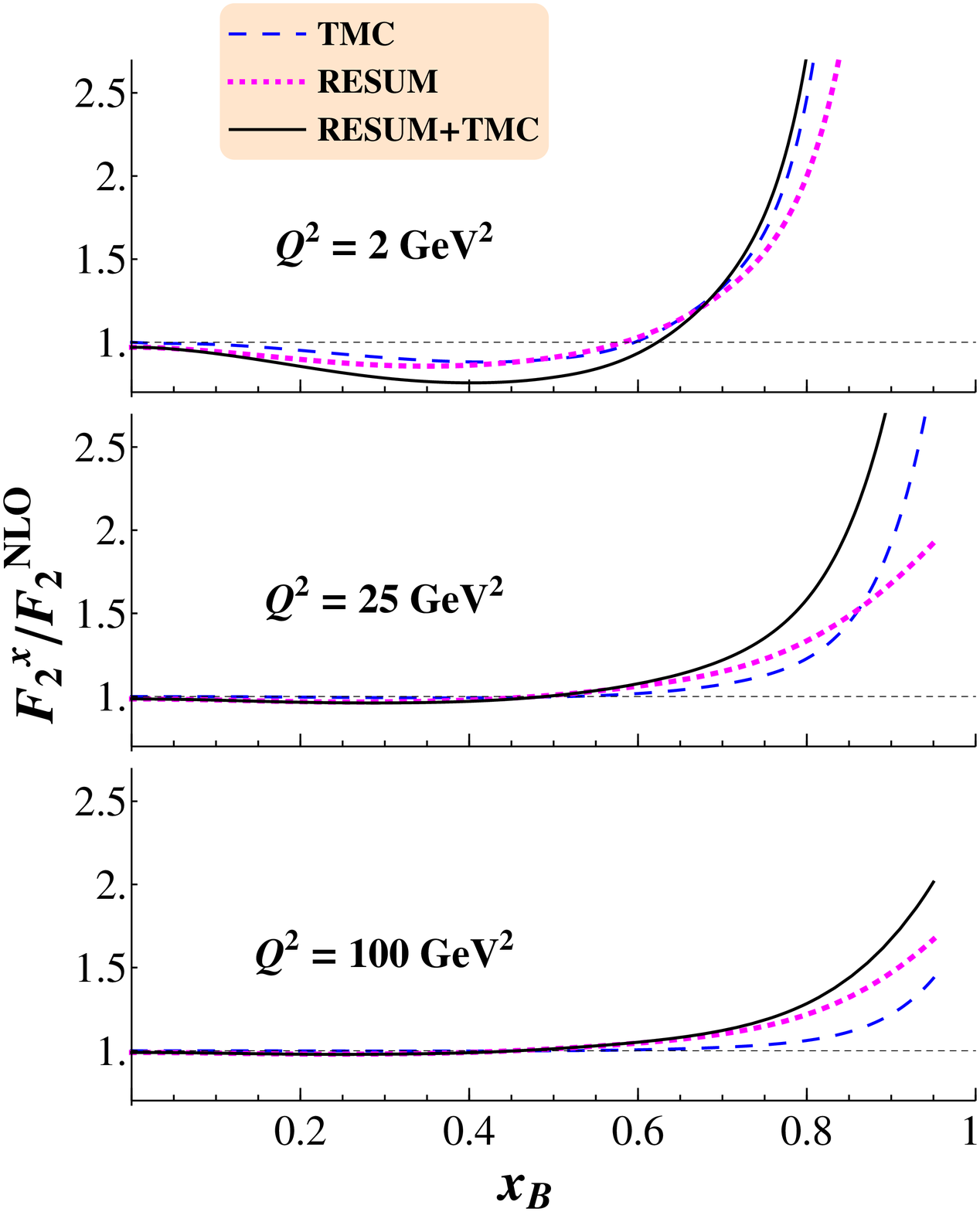}
\caption{\sf We show the effects of TMC (dashed blue), threshold resummation (dotted magenta) and the combination of both (solid black) normalized to NLO for the DIS structure function $F_2$ for different values of $Q^2=2,\,25,\,100~\mathrm{GeV}^2$. The PDF set of~\cite{Martin:2009iq} is used. \label{fig:DIStmcresueffects}}
\end{figure}

In order to get a deeper insight into the effects of TMCs on resummation, we can more closely analyze the integration region, that we depict in Fig.~\ref{fig:integration1} for a small and a large value of $Q^2$. The partonic threshold for resummation is set by $x=\xi$, as indicated by the blue dots. Hence, we may view the effect of TMCs as cutting out the singularities lying at  $x > \xi_{\mathrm{th}}$. As $Q^2$ increases, the amount of excluded singularities decreases, as can be seen from the diagram on the right. In the Bjorken limit ($Q^2 \rightarrow \infty$), $\xi_{\mathrm{th}}$ tends to 1, the integration region spans the whole triangle, and no singularity is excluded. Since the threshold for gluon radiation is set for $x\rightarrow \xi$ the threshold singularities appear only at the lower integration boundary of the first term, which is therefore the only one where large logarithms appear and need resummation. This can then be achieved without introducing a non-zero result for the resummed structure function in the unphysical region of $x_B>1$ because the first term in Eq.~\eqref{eq:DIS3} is zero at $x_B\geq 1$. In the second term, that also tends to zero as $x_B \rightarrow 1$, the threshold limit is not reached so that there is no need to regularize any of the terms in the coefficient function and we can treat this as part of the matching procedure to the full NLO calculation, see Eq.~\eqref{eq:matching}.

As in the massless target approximation, we derive threshold resummation in Mellin space but taking special care of the fact that at finite $Q^2$ the first term in Eq.~(\ref{eq:DIS3}) does not have the standard convolution structure as for the massless approximation of the structure functions. Taking Mellin moments with respect to $\xi$ of the first term in Eq.~(\ref{eq:DIS3}) only, we obtain
\begin{align}
\label{eq:mellinmoments}
{\cal F}^{\mathrm{TMC},N}_1 & 
  = \int_0^1d\xi\,\xi^{N-1} \int_{\xi}^{\xi_{\mathrm{th}}} \frac{dx}{x}
  {\cal C}^1_f\left(\frac{\xi}{x} \right)f(x)  \nn \\
& = \int_0^1d\xi\,\xi^{N-1} \int_0^1dy \int_0^{\xi_{\mathrm{th}}}dx\, 
  {\cal C}^1_f(y)\,f(x)\,\delta(xy-\xi)\nn \\
& = \left(\int_0^1dy\,y^{N-1}{\cal C}_f^1(y) \right) 
  \left(\int_0^{\xi_{\mathrm{th}}} dx\, x^{N-1}f(x)\right) \nn \\
& = {\cal C}_f^{1,N}\,f^N_{\xi_{\text{th}}} \, ,
\end{align}
where we denoted by $f^N_{X} = \int_0^X dx\, x^{N-1}f(x) $ the $N$-th truncated moment of a function $f$. Hence, In Mellin space, the TMC corrected structure function ${\cal F}^{\mathrm{TMC}}_1$ factorizes into a product of the moments of the coefficient function ${\cal C}_f^{1,N}$, exactly as in the massless approximation, and of the truncated moments of parton distributions. The appearance of the latter reflects the reduced support for integration over $x$ in Eq.~\eqref{eq:FTMC} (as illustrated in Fig.~\ref{fig:integration1}). The truncation of the PDF moments increases in magnitude with the increase of $x_B$ and the decrease of $Q^2$.

Using the resummed coefficient function ${\cal C}^{1,N}_{q,\mathrm{res}}$ in Eq.~(\ref{resummed1}), we may perform the inverse transformation,
\be\label{eq:mellininverse1}
{\cal F}^{\mathrm{TMC}}_{1,\mathrm{res}}(x_B,Q^2)=\int_{{\cal C}_N}\frac{dN}{2\pi i}\,\xi^{-N}\,{\cal C}^{1,N}_{q,\mathrm{res}}\, f^N_{\xi_{\mathrm{th}}}\, ,
\ee
using the same contour as in the massless target case, see e.g.~\cite{Anderle:2012rq}. Note that this corresponds only to resummation of the first term in Eq.~(\ref{eq:DIS3}). We always have to calculate the second term separately and add it to the resummed result. Other than that, the matching procedure required to include the full NLO calculation is the same as that without TMCs, see Eq.~(\ref{eq:matching}).

\subsection{Phenomenological Results\label{ssec:tmc3}} 

We now investigate the numerical effects of TMC and threshold resummation as well as their combination. Throughout this work we only consider a proton target. We make use of both the NLO ``Martin--Stirling--Thorne--Watt'' (MSTW 2008) set of parton distribution functions~\cite{Martin:2009iq} as well as the NLO CJ12 PDF set of~\cite{Owens:2012bv} . As shown in Eq.~(\ref{eq:mellinmoments}), in order to perform numerical calculations for threshold resummation, we have to compute Mellin moments of the PDFs. Since these are provided in $x$ space, we first fit suitable functions to the PDFs using the following parametrization
\be 
a_0\,x^{a_1}\, (1-x)^{a_2}\left(\sum_{j=3}^n a_j x^{b_j} \right),
\ee
where $a_i$ are free parameters and $b_j$ are some chosen fixed values in the range of $0$ to $3$. We take into account the $Q^2$ evolution of the PDFs by allowing a $Q^2$ dependence in the parameters $a_i$ of the form
\be
a_i=a_{i1}+a_{i2}\log(\log(Q^2/Q_0^2)) \, .
\ee 
The parameters $a_{i1,i2}$ are free parameters to be fitted for each different PDF and $Q_0^2$ is a chosen fixed scale. The truncated Mellin moments of the fitted PDFs are then taken analytically. With TMCs, we obtain a sum of incomplete Beta functions of the type
\be\label{eq:betafct1}
B_{{\xi_{\mathrm{th}}}}(N+a_1+b_j,a_2+1). 
\ee
The index $\xi_{\mathrm{th}}$ corresponds to the upper integration limit in the definition of the incomplete Beta function. (Without TMCs, or rather in the large $Q^2$ limit, where $\xi_\text{th}=1$, we obtain a sum regular Beta functions, $B_{1}$.)

\begin{figure}[t]
\vspace*{-2mm}
\hspace*{-.5cm}
\includegraphics[width=0.5\textwidth]{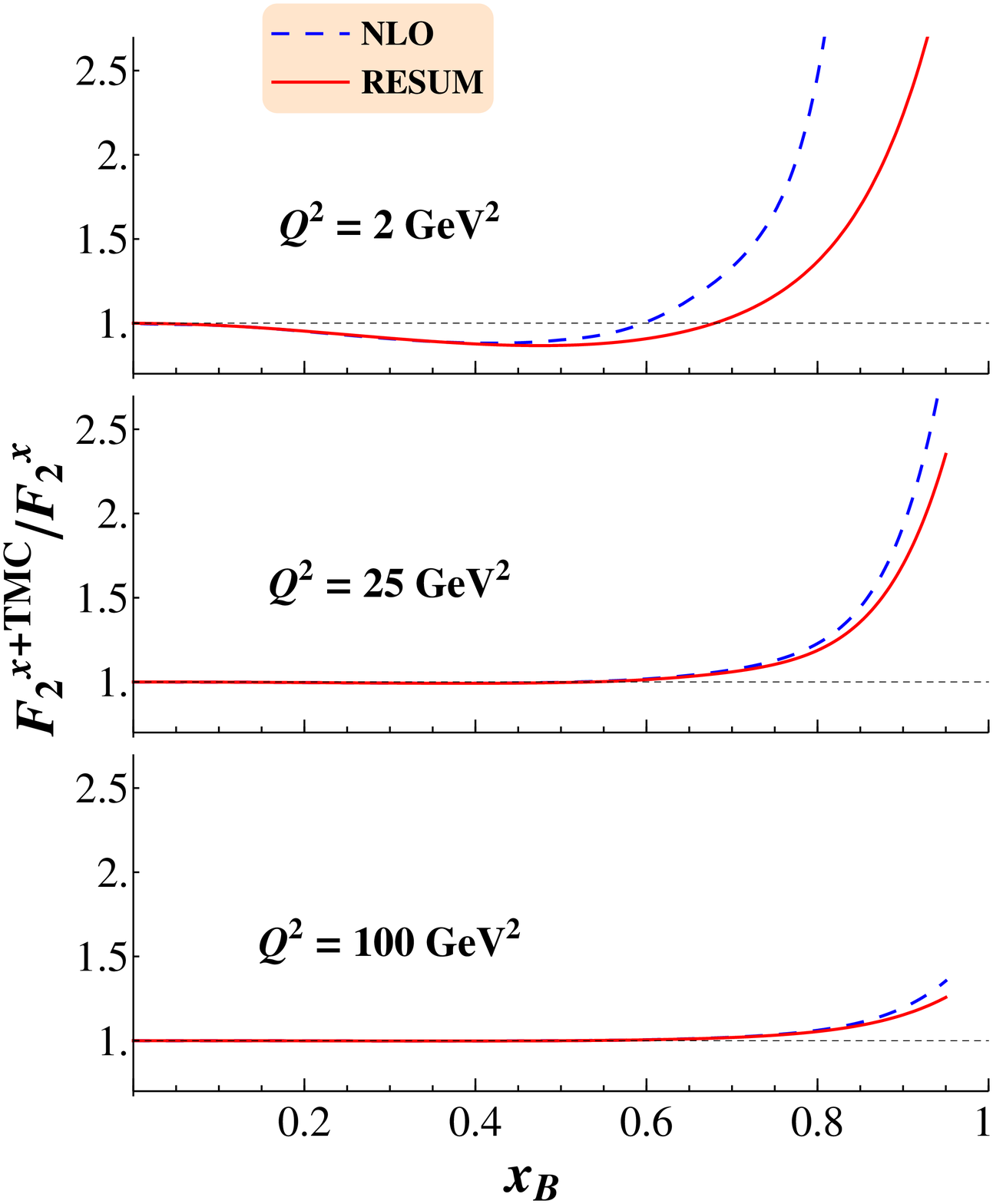}
\caption{\sf The effect of TMC is shown for the structure function $F_2$ on top of a NLO and a resummed calculation. We show TMC normalized to NLO (dashed blue) as well as TMC and resummation combined normalized to the resummed result (solid red). Again, we choose three representative values of $Q^2=2,\,25,\,100~\mathrm{GeV}^2$. The PDF set of~\cite{Martin:2009iq} is used.\label{fig:DIStmcresuinterplay}}
\end{figure}

In our code, we implement the incomplete Beta functions by making use of the identity
\ba\label{eq:betafct2}
&&\hspace{-.8cm}B_{\xi_{\mathrm{th}}}(N+a_1+b_j,a_2+1) = \frac{\xi_{\mathrm{th}}^{N+a_1+b_j}}{N+a_1+b_j}  \\
&&\hspace{-.3cm}\times \;{}_2F_1(N+a_1+b_j,-a_2,1+N+a_1+b_j,\xi_{\mathrm{th}}) \, , \nn
\ea
and for the complex hypergeometric function ${}_2F_1$ we use the routine provided in~\cite{Michel2008535}. In order to rule out uncertainties introduced in our calculation when using the fitted functions for the Mellin inversion, we checked the accuracy of the fits by comparing results at NLO obtained from the convolution code in $x$-space and the Mellin inverse. Indeed, we find very good agreement even for very large values of $x_B$. 

The reason behind the numerical stability of our result is the following.
When performing the Mellin inverse, we obtain an exponential suppression for large negative real values of $N$ due to the factor $\xi^{-N}$ in Eq.~(\ref{eq:mellininverse1}). 
When TMCs are included, this suppression is softened by the factor $\xi_{\mathrm{th}}^{N}$ in Eq.~(\ref{eq:betafct2}). These two exponential factors originate from two different parts of the calculation: the first comes from the definition of the inverse Mellin transform, whereas the second is due to the incomplete beta function. We need to combine the two factors into one single exponential, $\exp[-N\ln(\xi/\xi_{\mathrm{th}})]$, where the cancellations between the two is made explicit and makes the numerical integration over $dN$ well-behaved even for very large values of $x_B$.\\

In Fig.~\ref{fig:DIStmcresueffects}, we present our numerical results for the DIS structure function $F_2$ using the PDF set of~\cite{Martin:2009iq}. All results are normalized to the massless NLO calculation. We choose to plot our results only up to $x_B=0.95$ as non-perturbative effects are expected to set in for too large values of $x_B$, which is beyond the scope of this work. The two effects under consideration are shown separately in dashed blue (TMCs) and in dotted magenta (threshold resummation) for three representative values of $Q^2=2,\,25,\,100~\mathrm{GeV}^2$. Both TMC and resummation effects become increasingly large as $x_B$ tends to 1, as it is clear from the kinematic analysis presented in Sec.~\ref{sec:tmc}. Both vanish at small $x_B$, the former because the Nachtmann variable $\xi$ and the kinematic factor $\rho$ tend to their massless value of $x_B$ and 1, respectively, and the latter because the integrals are evaluated more and more far from the resummation threshold. Concerning the $Q^2$ dependence of the two corrections under discussion, both effects taken separately are large at small values of squared momentum transfer, and decrease with increasing $Q^2$. However, TMCs exhibit a power law suppression in $Q^2$, while resummation corrections decrease much less rapidly and become dominant, and non-negligible, at $Q^2 \gtrsim 25$ GeV. The results we find for TMCs are in agreement with numerical results in previous work such as~\cite{Brady:2011uy} and~\cite{Accardi2009} up to some prefactor conventions. Concerning the validity of our results on DIS threshold resummation one may compare to Ref. \cite{Sterman:2000pu,Anderle:2012rq}.

\begin{figure}[t]
\vspace*{-2mm}
\hspace*{0cm}
\includegraphics[width=0.52\textwidth]{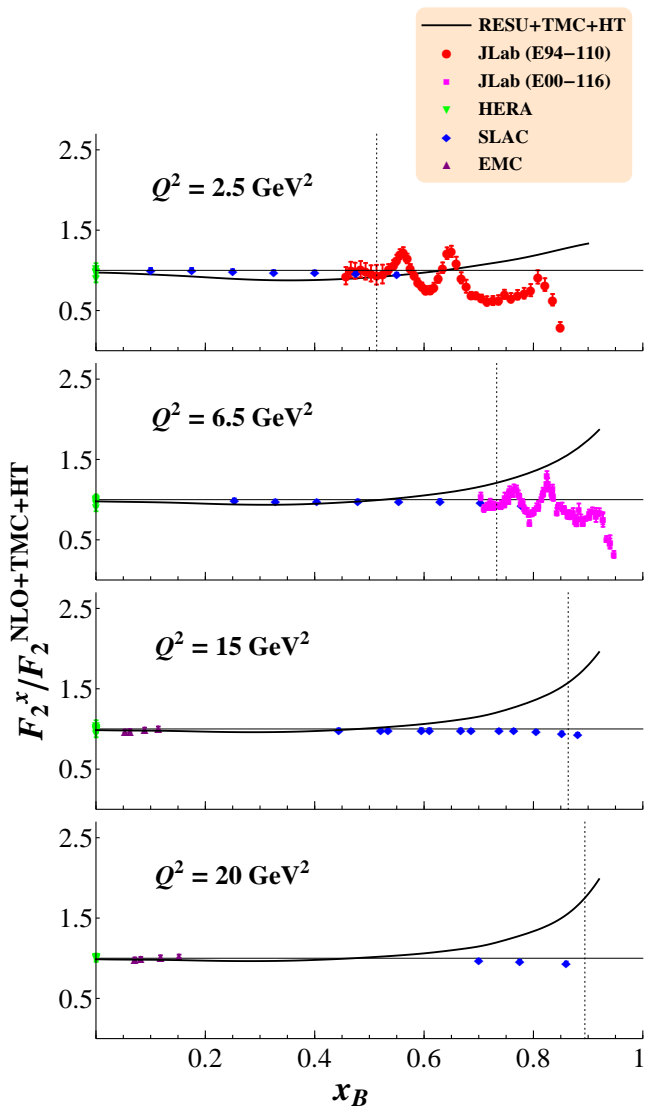}
\caption{\sf We plot ratios of data/theory for DIS structure function $F_2$ at several $Q^2$. Here ``theory'' denotes the NLO results with TMC and higher twist contributions based on the CJ PDF set of~\cite{Owens:2012bv}. The data is taken from~\cite{Liang:2004tj,Malace:2009kw,Aaron:2009aa,Whitlow:1991uw,Aubert:1985fx}. Due to our choice of a linear scale for the horizontal axis, the HERA data appears clustered at the vertical axis, i.e. at very small $x_B$. In addition, using the same normalization, we plot the theoretical prediction when resummation is included as well. The dotted line corresponds to $W^2=3.25\,\mathrm{GeV}^2$. \label{fig:disdata}}
\end{figure}

We can now turn to the combination of TMC and threshold resummation, shown by the solid black line in Fig.~\ref{fig:DIStmcresueffects}. We notice that the strength of the two effects does not add in a simple way. In order to understand the interplay of TMCs and threshold resummation, we analyze the plots in Fig.~\ref{fig:DIStmcresuinterplay}. Again, we use the PDF set of~\cite{Martin:2009iq}. There, we compare the ratio of the target mass corrected $F_2$ structure function to the massless calculation without resummation (Dashed blue line), and the ratio of the structure function with both TMC and resummation, but normalized to the resummed result (solid red line). This way, we can see how the TMC contribution acts on top of a purely NLO calculation compared to being added to a resummed calculation. Firstly, we note that the effects remain decoupled for small values of $x_B$, where both ratios lie exactly on top of each other. This decoupled region extends to larger values of $x_B$ as $Q^2$ increases. However, at large enough values of $x_B$ the two functions deviate and TMC acts differently for NLO than for the resummed result. 

As discussed in \cite{Accardi:2009br,Accardi:2013pra}, such a variation in the calculation of the $F_2$ structure function can lead to considerable difference in the value of the $d$-quark parton distribution extracted in a global fit. The theoretical description of the data crucially depends on whether resummation is included or not. In order to gauge the relevance of TMCs and resummation for the extraction of PDFs, but leaving a detailed QCD fit for future work, we present in Fig.~\ref{fig:disdata} a comparison of our calculations to a variety of electron-proton scattering data from  JLab (E94-110)~\cite{Liang:2004tj}, JLab (E00-116)~\cite{Malace:2009kw}, HERA~\cite{Aaron:2009aa}, SLAC~\cite{Whitlow:1991uw}, and EMC~\cite{Aubert:1985fx}. Here we use the CJ12 PDF set of~\cite{Owens:2012bv}. The data was bin-centered in $Q^2$ for the analysis of Nachtmann moments of the DIS longitudinal structure function in~\cite{Monaghan:2012et} allowing a direct comparison of different experimental results~\cite{note1}. The data was normalized to a calculation including TMCs only; but in order to do so we also need to add the ``residual'' power corrections in $1/Q^2$ not taken care of by target mass corrections. These were included in the CJ12 QCD fit \cite{Owens:2012bv} via a multiplicative factor $1+C(x_B)/Q^2$, with $C$ a parametrized function of Bjorken $x_B$ with parameters fitted to a variety of DIS data. We include the same multiplicative factor in our NLO calculation, and use the parameters obtained in the CJ12 fit. The vertical dotted line in Fig.~\ref{fig:disdata} corresponds to a value of $W^2=(P_h+q)^2=m_N^2+Q^2(1-x_B)/x_B=3.25\,\mathrm{GeV}^2$, which is generally regarded as the end of the DIS regime and the beginning of the resonance region where fluctuations of the data around the DIS calculation are generally understood in terms of quark-hadron duality \cite{Melnitchouk:2005zr}.

Finally, in order to gauge the relevance of resummation corrections to a global fit of parton distributions, we also plot in Fig.~\ref{fig:disdata} the structure function $F_2$ with resummation, TMCs and higher twist contributions, normalized by the pure NLO calculation including TMCs and higher twists which was also used to normalize the data. Comparing the obtained deviation of this curve from one with the experimental uncertainties, we find a very significant effect which is getting larger for increasing $Q^2$, while at low $Q^2$, TMCs already capture the main effects. In fact, threshold resummation also decreases with increasing $Q^2$, as can be seen from both Figs.~\ref{fig:DIStmcresueffects} and \ref{fig:DIStmcresuinterplay} above. However, as already remarked, TMCs die off rather quickly, whereas resummation remains clearly non-negligible in both the DIS and the resonance regions. Hence, resummation is likely to affect the extraction of large-$x$ partons (quarks directly, and gluons indirectly through QCD evolution in DIS) in global PDF fits. In this respect, it is important to remark that the non power law dependence of the resummation corrections cannot be effectively included in a phenomenological higher-twist term, and needs to be instead explicitly calculated in order to obtain the correct behavior of the quark PDFs at large values of the parton momentum fraction $x$. In particular it would be interesting to see how the effect is on the $u-$ and $d$-quark PDFs, and how much the extrapolation of the $d/u$ quark ratio to $x \rightarrow 1$ obtained in Ref.~\cite{Owens:2012bv} would be affected. \\ Finally, see also the work of~\cite{Blumlein:1998nv,*Christy:2012tk} concerning TMC effects for (polarized) structure functions.

\section{Hadron Mass Corrections and Resummation for SIA \label{sec:hmc}}

\subsection{Hadron Mass Corrections \label{ssec:hmc1}}

\subsubsection{Hadron level and parton level kinematics}

\begin{figure}[t]
\vspace*{-2mm}
\hspace*{.5cm}
\includegraphics[width=0.35\textwidth]{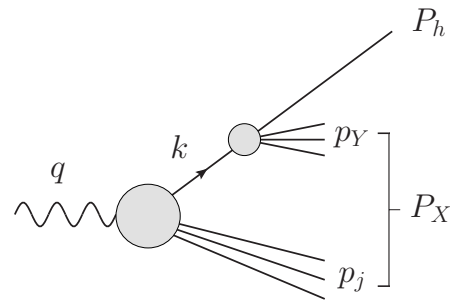}
\caption{\sf Diagram for SIA $e^+e^-\rightarrow hX$ where all momenta are specified. \label{fig:e+e-}}
\end{figure}

We study the kinematics for Single Inclusive electron-positron Annihilation hadron in the $\gamma\hspace{-.7mm}-\hspace{-.7mm} h$ frame, where both the photon $\gamma$ and the observed hadron $h$ have no transverse momentum component. We start by parametrizing the momenta of the virtual photon $q$, the observed hadron in the final state $P_h$ and the momentum of the fragmenting parton $k$. All momenta are also shown in Fig.~\ref{fig:e+e-}. We find,
\ba\label{eq:momenta1}
q^\mu & = & q^+\bar n^\mu+\frac{Q^2}{2q^+} n^\mu \, , \nn \\
P_h^\mu & = & \xi_E q^+\bar n^\mu+\frac{m_h^2}{2\xi_Eq^+}n^\mu \, ,\nn\\ 
k^\mu & = & \frac{\xi_E}{z} q^+\bar n^\mu+\frac{(k^2+\vect{k}_T^2)z}{2\xi_Eq^+} n^\mu
+\vect{k}_T\, ,
\ea
where $Q^2 = q^\mu q_\mu$ denotes the virtuality of the photon, $m_h$ is the mass of the observed hadron $h$, and $\xi_E = P_h^+/q^+$ its light cone momentum fraction; analogously, $z=P_h^+/k^+$ is the light-cone fractional momentum of the hadron relative to the parton that it is fragmenting from. 

The external Lorenz invariants are
\be
Q^2=q^2=s,\quad x_E=\frac{2q\cdot P_h}{q^2},\quad P_h^2=m_h^2 \, ,
\ee
where $s$ is the center of mass energy of the process. Solving for the virtual boson fractional momentum, we obtain
\be\label{eq:xi1}
\xi_E=\frac{P_h^+}{q^+}=\frac{1}{2}x_E\left(1+\sqrt{1-\frac{4}{x_E^2}\frac{m_h^2}{Q^2}} \right) \ , 
\ee
which is a ``Nachtmann-type'' fragmentation variable, {\em cf.} Eq.~\eqref{eq:Nacht}. Note that the radicand is always positive due to energy conservation at the hadron level, as we derive below. Inverting Eq.~(\ref{eq:xi1}) we obtain
\be\label{eq:xi2}
x_E=\xi_E\left(1+\frac{m_h^2}{\xi_E^2Q^2} \right) \, .
\ee
Concerning the unobserved parton's (internal) kinematics, we work in collinear factorization but refrain from fixing the value of the parton virtuality $k^2$ until we analyze the effects of non-zero hadron masses on the partonic kinematic bounds. Therefore, for the time being, we only set 
\be\label{eq:zandk}
\vect{k}_T=0 \, .
\ee 
Finally, we define the partonic fragmentation invariant $\hat x_E$ by
\be\label{eq:xipart}
\hat x_E=\frac{2k\cdot q}{q^2}=\frac{\xi_E}{z} +\f{zk^2}{\xi_EQ^2}\, ,
\ee
where the parton virtuality $k^2$ appears explicitly for the time being.

\subsubsection{Four momentum conservation and kinematic bounds}

\begin{figure}[b]
\hspace*{-.5cm}
\includegraphics[width=0.45\textwidth]{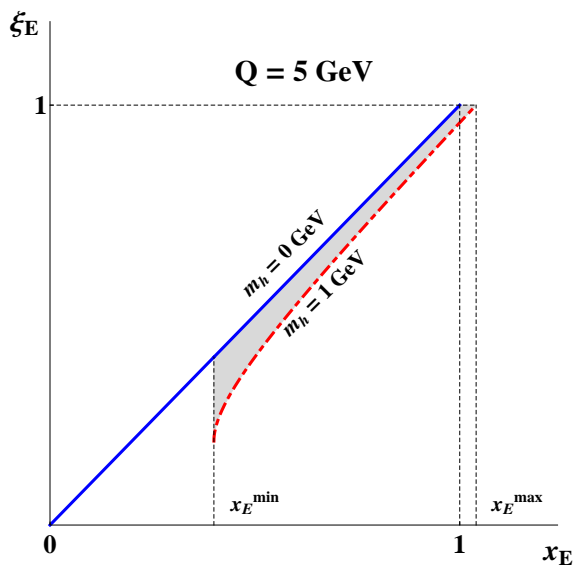}
\caption{\sf The fragmentation ``Nachtmann-type'' fragmentation variable $\xi_E$ as a function of $x_E$ at fixed $Q=5$ GeV. For illustration purposes, we choose a fictional mass of $m_h=1$ GeV (dash-dotted red) and compare it with a massless hadron, $m_h=0$ GeV (solid blue line). \label{fig:xiExE}}
\end{figure}

We consider now the kinematics at the hadron level, and derive the kinematic limits for $x_E$ and $\xi_E$ due to four momentum conservation. Firstly, we find a lower bound for $x_E$ which ensures that $\xi_E$ in Eq.~(\ref{eq:xi1}) is well defined. Calculating in the $e^+e^-$ c.m. frame with $q^+=q^-=Q/\sqrt{2}$, we find
\ba\label{eq:xEmin}
x_E & = & \f{2P_h\cdot q}{Q}=\f{\sqrt{2}}{Q}(P_h^++P_h^-)=\f{2E_h}{Q} \nn \\
& \geq & \f{2m_h}{Q} \equiv x_E^{\mathrm{min}}\, .
\ea
As a next step, we may derive an upper bound by considering the overall momentum conservation at the hadron level, $q=P_h+P_X$. We find
\begin{align}
0\leq P_X^2 & =  (P_h-q)^2 =  m_h^2 - x_E Q^2 + Q^2 \, .
\end{align}
Hence, 
\begin{align}
  x_E\leq 1+m_h^2/Q^2\equiv x_E^{\mathrm{max}} \, ,
\end{align}
which implies that $x_E$ can become slightly larger than one. This is due to the neglect of hadron mass effects in the unobserved hadron jet shown at the bottom of Fig.~\ref{fig:e+e-}. This is in analogy to what we did when analyzing the DIS kinematics. Using these two relations, we may determine the minimal and maximal values for $\xi_E$, which are
\ba
&& \xi_E^{\mathrm{min}}=\f{x_E^{\mathrm{min}}}{2}=\f{m_h}{Q} \nn \\
&& \xi_E^{\mathrm{max}} = 1\, .
\ea
With these limits at hand, we may plot $\xi_E$ as a function of $x_E$, see Fig.~\ref{fig:xiExE}. Here, the effects of hadron mass corrections are large when the invariant $x_E$ is small, contrary to the case of DIS, where target mass corrections are most relevant at large values of $x_B$. This can be understood as a consequence of crossing symmetry on the kinematics of the process, where now the virtual photon is time-like.

In a second step, again analogously to the procedure for DIS, we analyze the kinematics at the parton level.
Firstly, we consider the hard-scattering vertex which corresponds to the lower grey circle in Fig.~\ref{fig:e+e-}. Using momentum conservation at the vertex $q=k+p_j$ and neglecting any non-zero lower bound for the mass of the recoiling jet, we obtain the following constraint
\begin{align}\label{eq:ineq1}
  0\leq p_j^2  = (q-k)^2 
  = Q^2 \left(1-\frac{z}{\xi_E}\frac{k^2}{Q^2}\right) \frac{z-\xi_E}{z} \, .
\end{align}
Secondly, we consider the hadronization vertex which corresponds to the upper right grey circle in Fig.~\ref{fig:e+e-}, and we apply again four-momentum conservation. We obtain
\begin{align}\label{eq:ineq2}
  0\leq p_Y^2 =  (k-P_h)^2
  = \left( z k^2 - m_h^2 \right) \frac{1 - z}{z}\, .
\end{align}
Interestingly, in SIA there appears no ``threshold problem'', as is the case in DIS, and $\xi_E$ can range all the way up to 1. This is due to the system having no net baryon number, contrary to the case of DIS where the net baryon number is 1, and its conservation needs to be explicitly taken into account in the parton-level kinematics. It is also important to notice that while the parton virtuality $k^2$ in the first of these inequalities is parametrically suppressed at large $Q^2$, no hard scale suppresses this nor the hadron mass $m_h$ in the second inequality. Therefore, it is not possible to define a ``massless hadron limit'' as was done for the DIS case, where the nucleon mass, $m_N$, always appears divided by $Q$. 
The physical solutions of Eqs.~\eqref{eq:ineq1}-\eqref{eq:ineq2} are:
\begin{gather}
  \xi_E \leq z \leq 1 \label{eq:zlimits}\\
  m_h^2 \leq z k^2 \leq \xi_E Q^2 \ .
\end{gather}
In particular, the quark virtuality must always be larger than $m_h^2$ because this value corresponds to the minimum invariant mass of the parton fragmentation products when a hadron of flavor $h$ is detected. Following our philosophy, we should then perform the collinear expansion  around an on-shell massive quark rather than around $k^2 = 0$. However, dealing with the subtleties involved in proving the factorization theorem at NLO for this case goes beyond the scope of this paper and is left for future work~\cite{Accardi-Anderle-Ringer-2}. Here, instead, we use the well known collinear factorization theorem for massless, $k^2=0$, fragmenting partons as in \cite{Albino:2008fy,Albino:2005gd} and we continue to explore the interplay of hadron mass corrections and threshold resummation.

\subsubsection{Cross section at NLO}

In order to compare our results to the SIA measurements from BELLE and BaBar, we need to compute hadron multiplicities in $e^+e^-\rightarrow hX$ which are defined as
\be
R_{e^+e^-}^h \equiv \frac{1}{\sigma^{\mathrm{tot}}}\frac{d^2\sigma^h}{dx_Ed\cos\theta}\, .
\ee
Here the hadron $h$ is produced at an angle $\theta$ relative to the initial positron. $\sigma^{\mathrm{tot}}$ denotes the totally inclusive cross section for $e^+e^-\rightarrow X$. At NLO, this is given by
\be
\sigma^{\mathrm{tot}} = \frac{4\pi\alpha^2}{3Q^2}N_c\sum_q e_q^2\left(1+\frac{\as}{\pi}\right) \, ,
\ee
where $N_c=3$ is the number of colors and $\alpha$ is the electromagnetic fine structure constant. As mentioned before, we may write the differential cross section $d^2\sigma^h/dx_Ed\cos\theta$ in terms of two structure functions which we denote as $\hat{\cal F}^h_i$ (i=1,L), cf.~\cite{Nason:1993xx,Furmanski:1981cw,Anderle:2013pla}. Including HMC, we find
\ba
\label{eq:epem}
\frac{d^2\sigma^h}{dx_E d\cos\theta} &=& \frac{\pi\alpha^2}{Q^2} N_c \f{1}{1-\f{m_h^2}{\xi_E^2Q^2}}\left[ \frac{1+\cos^2\theta}{2} \hat{{\cal F}}^h_1(x_E,Q^2)\right.\nn\\[2mm]
&&+ \sin^2\theta\, \hat{{\cal F}}^h_L(x_E,Q^2) \Big]\nn \\
& = & \left. \f{1}{1-\f{m_h^2}{\xi_E^2Q^2}} \frac{d^2\sigma^h}{d\xi_E d\cos\theta}\right|_{x_E=\xi_E}  \, ,
\ea
where the Jacobian factor of $1/(1-m_h^2/\xi_E^2Q^2)$ is included in order to obtain a cross section differential in $x_E$ instead of $\xi_E$ \cite{Albino:2008fy,Albino:2005gd}. The structure functions $\hat {\cal F}_i^h$ with HMCs take into account the kinematic bounds on $z$ from Eq.~\eqref{eq:zlimits} and read
\beq
\label{eq:f1epem}
\hat{{\cal F}}_i^h(x_E,Q^2) =\sum_f \int_{\xi_E}^1 \frac{dz}{z} D_f^h \left(z,\mu^2\right) 
\hat{{\cal{C}}}^i_f \left(\frac{\xi_E}{z},\frac{Q^2}{\mu^2},\alpha_s(\mu^2)\right)\, ,
\eeq
where $D_f^h(z,\mu^2)$ denotes the fragmentation function for an observed hadron $h$ in the final state resulting from a parent parton $f$. The $\hat{\cal C}_f^i$ are the corresponding coefficient functions which we list in Appendix~\ref{ap:SIAcoefFunc} for completeness up to NLO. The cross section without HMCs is obtained by replacing $\xi_E$ with $x_E$ in Eq.~(\ref{eq:f1epem}) and by setting $m_h=0$ in Eq.~(\ref{eq:epem}). Having chosen to factorize the cross section around a parton virtuality $k^2=0$ this massless hadron limit can also be achieved in the $Q^2 \rightarrow \infty$ limit. 

\subsection{Combining HMC and Threshold Resummation \label{ssec:hmc2}}

\begin{figure}[hb!]
\hspace*{-.5cm}
\includegraphics[width=0.45\textwidth]{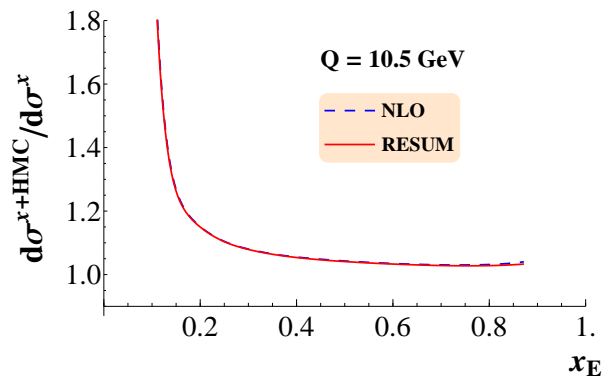}
\caption{\sf Comparison of the effect of HMC on top of NLO (dashed blue) and the resummed result (solid red) for $Q=\sqrt{s}=10.5$ GeV and the kaon mass $m_{K^0}=497.6$~MeV.  \label{fig:HMCcompare}}
\end{figure}

In the spirit of ``crossed resummation''~\cite{Sterman:2006hu}, we note that the only difference concerning the resummation in SIA in comparison to DIS is that we have to adjust one term in the matching coefficient $H_q$ in Eq.~(\ref{Hq}) $-\pi^2/6\rightarrow 5\pi^2/6$, see also~\cite{Anderle:2012rq,Cacciari:2001cw}. This similarity may be understood in the sense that both processes have one ``observed'' and one ``unobserved'' parton. Hence, the threshold resummed expression may again be written as a product of the form $H_q'\,\Delta^N_q\,J^N_q$. HMC and resummation are combined by simply replacing $x_E\rightarrow \xi_E$ in the resummed formula. There are no issues with $\xi_{\mathrm{th}}$ as it was the case for DIS, since the upper integration limit for $z$ in Eqs.~(\ref{eq:zlimits}),~(\ref{eq:f1epem}) is left unchanged compared to the massless hadron calculation. Since resummation effects increase with $x_E$ and HMC effects become large at small values of $x_E$, we do not expect a significant interplay of the two, contrary to the DIS case in which both effects increase at large $x_B$. We can numerically assess the interplay of HMC and threshold resummation similarly to what we did for DIS. In Fig.~\ref{fig:HMCcompare}, we plot the cross section including the effect of HMCs on top of an NLO (dashed-dotted blue line) and a resummed (solid red line) calculation. These are normalized to the corresponding massless hadron calculation to highlight HMC effects. We find that both ratios match completely. Hence, there is no crosstalk between the two effects.

\subsection{Phenomenological Results \label{ssec:hmc3}} 

Given the actuality of the recent BELLE~\cite{Leitgab:2013qh} and BaBar~\cite{Lees:2013rqd} results, we choose to present our numerical results for HMC and threshold resummation directly in comparison to data. The BELLE experiment is operating at a c.m.s. energy of $\sqrt{s}= 10.52$~GeV and similarly BaBar at $\sqrt{s}=10.54$~GeV, just below the lower end of the energy range of experiments typically included in FF fits, see for example Refs.~\cite{Albino:2008fy,deFlorian:2007aj}. 
This way, we maximize the effect of HMC and resummation of threshold logarithms and we may directly evaluate the significance of the two corrections compared to statistical and systematic uncertainties of the data. 

For the plots we discuss in this section, as well as for that in Fig.~\ref{fig:HMCcompare}, we used the ``de Florian-Sassot-Stratmann''~\cite{deFlorian:2007aj} set of fragmentation functions at NLO, where the new data from BELLE and BaBar is not yet included.
The goal is to show the phenomenological importance of threshold resummation and HMCs, and to qualitatively assess their relevance in global FF fits, rather than obtain a perfect description of the data. Comparing the size of HMC and threshold resummation to statistical and systematical errors, we will conclude that a fit including the two effects may yield rather different results for the extracted FFs. Whether indeed a better $\chi^2$ can be obtained given all the other data sets used in a global fit, as the study presented in 
\cite{Albino:2008fy} indicates, will be left for future work. 

Both BELLE and BaBar have an angular coverage of $-1<\cos\theta<1$. Hence, we integrate over the full range of $\cos\theta$ and obtain a cross section differential only in $x_E$. An important difference between the two data sets is that BELLE data is presented as a function of the Lorentz invariant energy fraction $x_E$, whereas BaBar is using the momentum fraction variable
\be
x_p=\f{2|\vect{p}_h|}{\sqrt{s}}\, .
\ee
Only for massless calculations are these equivalent, however, and in particular for kaons at present energies the difference between $x_E$ and $x_p$ is quite significant. When comparing our results to data, we multiply the BaBar data set by $J = dx_p/dx_E$ to obtain a cross section differential in $x_E$ and compare this to measurements at BELLE.

\begin{figure*}[t]
\centering
\includegraphics[width=\textwidth,clip]{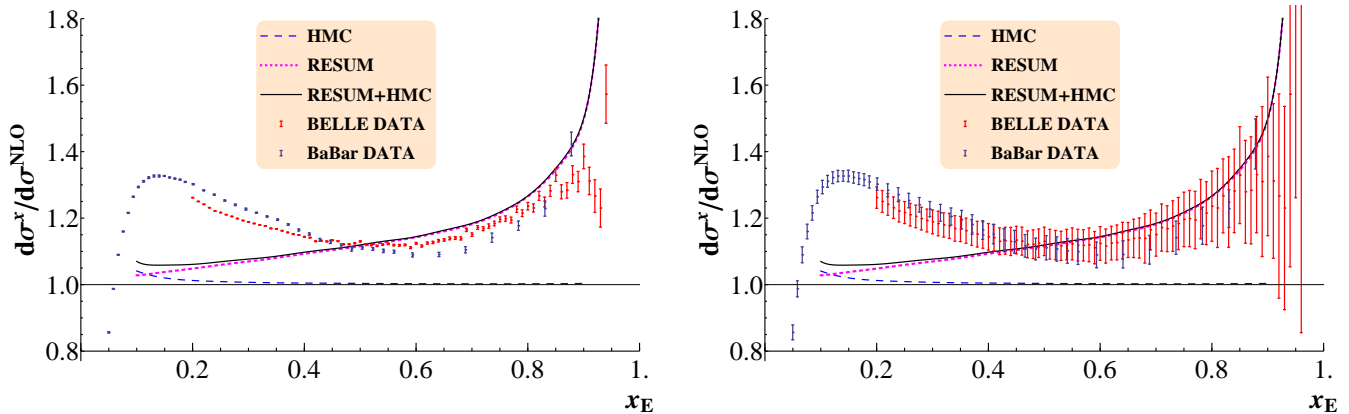}
\caption{\sf Both theory and data $1/\sigma_{\mathrm{tot}}\,d\sigma/dx_E$ are normalized to NLO for charge integrated pions at $\sqrt{s}=10.5$~GeV. The dashed blue lines shows the HMC corrected multiplicities, magenta dotted the resummed calculation and solid black the combination of both. BELLE data (red) and BaBar data (blue) is shown along with statistical (left) and systematical uncertainties (right). The FFs of~\cite{deFlorian:2007aj} are used. \label{fig:epempion}}
\end{figure*}

\begin{figure*}[t]
\centering
\includegraphics[width=\textwidth,clip]{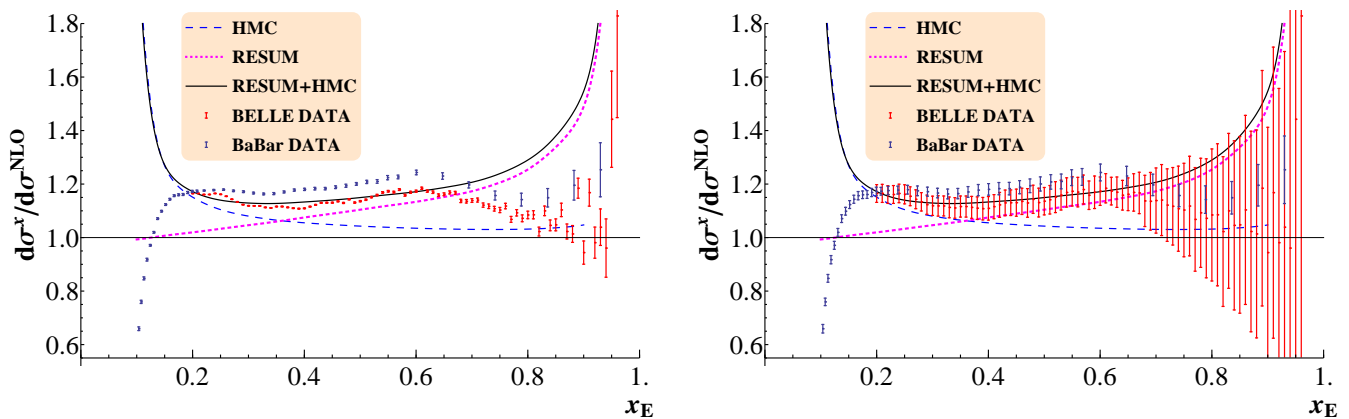}
\caption{\sf Same as Fig.~\ref{fig:epempion} but for observed kaons. \label{fig:epemkaon}}
\end{figure*}

We start by analyzing our calculations for (charge integrated) pion production, plotted in Fig.~\ref{fig:epempion}. All multiplicities $1/\sigma_{\mathrm{tot}}\,d\sigma/dx_E$ presented here are normalized to the calculation at NLO without hadron mass corrections. Our results for HMC (dashed blue line), threshold resummation (dotted magenta line) and the combination of both (solid black line) is shown. On the left (right) panels of Fig.~\ref{fig:epempion}, we show BELLE and BaBar data with statistical (systematic) uncertainties. As expected, the effects of threshold resummation are quite significant and most relevant at large $x_E$, whereas HMCs affect the calculation at small $x_E$, and in the measured range are not large, due to the smallness of the pion mass, $m_{\pi^0}=135$~MeV. Nonetheless, given the statistical precision of this data, it seems important to account for HMCs in a global fit. Much of discrepancy between NLO calculations and pion data can be resolved by including the new data in a global FF fit, as it was very recently shown in Ref.~\cite{deFlorian:2014xna}. Finally, we note that below $x_E=0.1$, small-$x$ logarithms start to become relevant and would also need to be resummed \cite{Albino:2005gd}, which however is beyond the scope of this work. 

For kaons, with mass $m_{K^0}=497.6$~MeV, HMCs are much larger than for pions, as shown in Fig.~\ref{fig:epemkaon}. The combination of HMC and resummation leads to a significant increase of the cross section compared to a massless hadron NLO calculation for all values of $x_E$, and their inclusion in global FF fits is even more important than in the pion case. 
The steep rise of the HMC corrected result over the NLO calculation at small $x_E$ is mostly due to the kinematic limit $x_E> 2m_{K^0}/Q \approx 0.1 $ derived in Eq.~\eqref{eq:xEmin}, and in its vicinity the validity of our treatment of HMCs may come into question. This is also the region where resummation of small-$x$ logarithms becomes important, and a proper treatment of these is likely to require a careful consideration of the interplay with HMCs. It would then be very interesting to explore the similarities and differences of this with the interplay of threshold logarithms and target mass corrections in large-$x_B$ DIS events we have discussed in Section~\ref{sec:tmc}, but we defer this analysis to a future effort.

\section{Conclusions \label{sec:conclusions}}

We have investigated two phenomenologically important effects for the analysis of data in inclusive DIS and single-inclusive electron-positron annihilation, namely the corrections to NLO calculations due to a non-zero mass of the nucleon target in DIS, and of the detected hadron in SIA, as well as the resummation of threshold logarithms arising in the perturbative expansion of the hard scattering coefficients. In both cases, these lead to a non-negligible enhancement in the calculated observable compared to the precision of the currently available experimental data. Therefore, both effects are significant for precise QCD fits of PDFs as well as FFs.

In DIS, target mass corrections and threshold resummation are both most relevant at large values of $x_B$. In particular, we have derived a way to perform resummation respecting the parton level kinematic constraints arising from consideration of the non-zero target mass. The resulting structure functions can then be consistently combined with TMC calculations such that they remain zero in the unphysical region $x_B\geq 1$.  We find that two effects are coupled especially for small values of $Q^2$. 
At large $x_B$, the size of the combined TMC and resummation corrections is considerably larger than the accuracy of the existing DIS data over an extended $Q^2$ range. Therefore, it should be taken into account for a precise extraction of large-$x$ PDFs in global fits.

In SIA processes, hadron mass corrections are relevant at small $x_E$ while threshold resummation is important at large $x_E$, and we find no interplay of the two effects. We have included both in our calculations of cross sections for pion and kaon production, and compared these to recent data from the BELLE and BaBar collaborations. The effects are again large, and non-negligible for the extraction of FFs, given the precision of the new data sets.
This is particularly true for kaons due to their bigger mass compared to observed pions. Given this large effect for kaon SIA, it becomes a topic of practical as well as theoretical interest to determine what the interplay is between the finite mass kinematics and the resummation of small-$x$ logarithms. We leave this for future efforts.

Finally, we remark that we have performed calculations in collinear factorization around massless, on-shell partons. For SIA, we have found that this choice, however commonly made, actually violates parton-level four momentum conservation. A detailed analysis of collinear factorization with non-zero virtuality partons is in preparation.

\begin{acknowledgments}
We are grateful to Werner Vogelsang for support and helpful discussions. We would like to thank Peter Monaghan, Marco Stratmann, and Christian Weiss for useful conversations. D.P.A. and F.R. acknowledge the kind hospitality of Hampton University and JLab, where part of this work was carried out. This work was supported by the German Research Foundation (grant VO 1049/1-1) and by the US Department of Energy grant No. DE-SC008791. In addition, D.P.A. was partially supported by a grant from Fondazione Cassa Rurale di Trento.
\end{acknowledgments}

\appendix
\section{DIS Coefficient Functions \label{ap:DIScoefFunc}}

The DIS coefficient functions up to NLO in the $\overline{\text{MS}}$ scheme are given by \cite{Gluck:1995yr,Furmanski:1981cw,deFlorian:2012wk}
\begin{align}\label{eq:coef_funcs}
{\cal C}_q^1(\hat{x}) = & e^2_q \delta(1-\hat{x})+e^2_q\f{\as}{2\pi}C_F\Bigg[(1+\hat{x}^2)\left(\f{\ln(1-\hat{x})}{1-\hat{x}}\right)_+  \nn \\
&-\f{3}{2}\f{1}{(1-\hat{x})}_+-\f{1+\hat{x}^2}{1-\hat{x}}\ln \hat{x}+3\nn\\
& -\left(-\f{9}{2}+\f{\pi^2}{3}\right)\delta(1-\hat{x}) \Bigg] \nn \\
{\cal C}_q^L(\hat{x}) = &e^2_q \f{\as}{2\pi}C_F\, 2\hat{x}\nn \\
{\cal C}_g^1(\hat{x}) = &e^2_q \f{\as}{2\pi}C_F\, \bigg[\left(\hat{x}^2+(1-\hat{x})^2\right)\ln\left(\f{1-\hat{x}}{\hat{x}}\right) \nn\\
& -1+4 \hat{x}(1-\hat{x})\bigg]\nn\\
{\cal C}_g^L(\hat{x}) = & e^2_q \f{\as}{2\pi}C_F\,[4 \hat{x} (1 - \hat{x})]
\end{align}
where $j=\,q,\,g$ and $\hat{x}=\xi/x$. The definitions of $x$ and $\xi$ are given in Eqs.~(\ref{eq:frac_mom}) and~(\ref{eq:Nacht}). These coefficients satisfy
\be
\label{eq:C2}
{\cal C}_j^2(\hat{x}) =  {\cal C}_j^1(\hat{x})+{\cal C}_j^L(\hat{x}) \, .
\ee

These coefficient functions are related to the coefficient functions $h_i$ ($i=1,2,L$) defined in~\cite{Accardi2009} as follows
\begin{gather}
\label{eq:rel_coefAccardi}
{\cal C}^1(\hat{x})=2 h_1(\hat{x}) \quad
{\cal C}^L(\hat{x})=2 h_L(\hat{x}) \quad
{\cal C}^2(\hat{x})=\f{h_2(\hat{x})}{\hat{x}}
\end{gather}
so that 
\be
h_L=-h_1+\f{h_2}{\hat{x}}. 
\ee
The structure function $F^{\text{AQ}}_{i}$ ($i=1,2,L$) presented in the same paper are connected to the ones defined in Eq.~(\ref{eq:Ftmc}) by the following relations
\ba
\label{eq:rel_sfuncsAccardi}
F^{\text{AQ}}_{1}(x_B,Q^2) & = & \f{{\cal F}^{\mathrm{TMC}}_1(x_B,Q^2)}{2}= F_1^{\mathrm{TMC}}(x_B,Q^2)\nn\\
F^{\text{AQ}}_{2}(x_B,Q^2) & = & x_B {\cal F}^{\mathrm{TMC}}_2(x_B,Q^2)= F_2^{\mathrm{TMC}}(x_B,Q^2)\nn\\[2mm]
F^{\text{AQ}}_{L,}(x_B,Q^2) & = & \f{{\cal F}^{\mathrm{TMC}}_L(x_B,Q^2)}{2}= \frac{ F^{\mathrm{TMC}}_L(x_B,Q^2)}{2 x_B}
\ea
so that 
\begin{equation}
F^{\text{AQ}}_{L}= \f{\rho^2}{2 x_B}F^{\text{AQ}}_2-F_1^{\text{AQ}},
\end{equation}
where $\rho$ is defined in Eq.~(\ref{eq:rho}).

\section{SIA Coefficient Functions \label{ap:SIAcoefFunc}}

The coefficient functions up to NLO for SIA in the $\overline{\text{MS}}$ scheme are given by \cite{Nason:1993xx,Furmanski:1981cw,Kretzer2000,Florian1997}
\begin{align}
\hat {\cal C}_q^1(\hat{z}) = & e^2_q\delta(1-\hat{z})+e^2_q\f{\as}{2\pi}C_F\left[(1+\hat{z}^2)\left(\f{\ln(1-\hat{z})}{1-\hat{z}}\right)_+ \right. \nn \\
&-\f{3}{2}\f{1}{(1-\hat{z})}_++2\f{1+\hat{z}^2}{1-\hat{z}}\ln \hat{z}+\f{3}{2}(1-\hat{z})\nn\\
&\left.+\left(\f{2}{3}\pi^2-\f{9}{2}\right)\delta(1-\hat{z})\right] \nn
\end{align}
\begin{align}\label{eq:epemcoef_funcs}
\hat {\cal C}_q^L(\hat{z}) = &e^2_q \f{\as}{2\pi}C_F\nn\\
\hat {\cal C}_g^1(\hat{z}) = &  e^2_q\f{\as}{2\pi} C_F\, 2\bigg[\f{1+(1-\hat{z})^2}{\hat{z}}\ln\left(\hat{z}^2(1-\hat{z})\right) \nn \\
& -2 \f{(1-\hat{z})}{\hat{z}}\bigg]\nn\\
\hat {\cal C}_g^L(\hat{z}) = & e^2_q \f{\as}{2\pi}C_F\,\left[4 \f{(1-\hat{z})}{\hat{z}}\right] \ ,
\end{align}
where $j=\,q,\,g$ and $\hat{z}=\xi_E/z$. The definitions of $z$ and $\xi_E$ are given in Eqs.~(\ref{eq:zandk}) and~(\ref{eq:xi1}). The listed coefficient functions are related by
\be
\label{eq:epemC2}
\hat {\cal C}_j^2(\hat{x}) =  \hat {\cal C}_j^1(\hat{x})+\hat {\cal C}_j^L(\hat{x}) \, .
\ee


	\bibliographystyle{apsrev}  
	\bibliography{library}

\begin{thebibliography}{74}
\bibitem[{\citenamefont{we~thank P. Monaghan for kindly providing us with the
  bin-centered~data points.}()}]{note1}
\bibinfo{author}{\bibnamefont{we~thank P. Monaghan for kindly providing us with
  the bin-centered~data points.}}
\end{thebibliography}

\end{document}